\documentclass[twocolumn,aps,prb,showpacs]{revtex4-1}
\usepackage[utf8]{inputenc}
\usepackage[normalem]{ulem}
\usepackage[T1]{fontenc}
\usepackage[unicode=true, pdfusetitle, bookmarks=true, bookmarksnumbered=false, bookmarksopen=false, breaklinks=false, pdfborder={0 0 1}, backref=false, colorlinks=false]{hyperref}
\hypersetup{colorlinks,linkcolor=blue,citecolor=blue,urlcolor=blue}
\usepackage{float}
\usepackage{verbatim}
\usepackage[abs]{overpic}
\usepackage{color}
\usepackage{comment}
\usepackage{amsmath,amssymb}
\usepackage{graphicx}
\usepackage{subcaption}

\begin{document}
\title[Example of an Article with a Long Title]
{What Is the Generalized Representation of Dirac Equation in Two Dimensions?}
\author{H. Moaiery}
\author{A. Chenani}
\author{A. Hakimifard}
\author{N. Tahmasebi}
\affiliation{Jundi-Shapur University of Technology, Dezful, Iran}

\begin{abstract}
	In this work, the general form of $2\times2$ Dirac matrices for 2+1 dimension is found. In order to find this general representation, all relations among the elements of the matrices and matrices themselves are found,and the generalized Lorentz transform matrix is also found under the effect of the general representation of Dirac matrices. As we know, the well known equation of Dirac, $ \left( i\gamma^{\mu}\partial_{\mu}-m\right) \Psi=0 $,
	is consist of matrices of even dimension known as the general representation of Dirac matrices or Dirac matrices. Our motivation for this study was lack of the general representation of these matrices despite the fact that more than nine decades have been passed since the discovery of this well known equation. Everyone has used a specific representation of this equation according to their need; such as the standard representation known as Dirac-Pauli Representation, Weyl Representation or Majorana representation. In this work, the general form which these matrices can have is found once for all.
\end{abstract}


\maketitle

\section{Introduction}

Quantum Mechanics and Special Relativity are considered as two major revolutions in twentieth-century physics.
The Dirac equation is undoubtedly one of the best examples of the link between these two fields. Although this equation was first proposed for the electrons, but it has performed great in describing muons, quarks, and tauons and its application in the fundamental and theoretical physics fields is obvious to
everyone, including Elementary particle, Quantum Chromodynamics, phenomenological
models of hadrons, in standard models, and even in cosmology~\cite{Ref1,Ref2,Ref3,Ref4}. In~\cite{Ref5} and ~\cite{Ref6}, it was demonstrated
that the Dirac equation not only can be applied to Fermionic states but also to Bosonic states.\\
In the last few decades, it has been demonstrated that electron in the shells closest to the
heavy nucleus influence the bonds of their included molecules in which the relativistic effects
are observable and some models such as Dirac-Hartree-Fock (DHF) are suggested to describe
the characterization of heavy elements. The use of relativistic methods is essential for
theoretical models in the quantum chemistry fields, including actinides, lanthanides, and
transition metals~\cite{Ref7}. The Dirac equation has many applications in the fields of electronic
transfers (ET)~\cite{Ref8}, Photonic Structures, and Supercooled Materials in optic networks~\cite{Ref9,Ref10,Ref11}.\\
Understanding the Dirac fermions is essential for the modern condensed matter
physics (CMP). There is a fundamental similarity among a wide range of materials such as d-wave
superconductors, graphene, and topological insulators so that their low energy
fermionic excitations acts similar to massless Dirac fermions. The Dirac fermions specific
behaviors are integrated into a unified framework for a class of materials named "Dirac
materials" in the condensed matter systems since many of the special properties of the
mentioned materials are induced by the Dirac spectrum of their particles. In other words, the
materials include low energy excitations with Dirac points involving linear dispersion
relation. In the absence of a mass term, all of the materials are gap-free. Dirac materials
create a different class of material against, for example, metals and semiconductors. Different
types of Dirac materials have been discovered so far, from normal state crystal materials to
asymmetric quantum fluids such as Si, Na3Bi, Cd3As2 ~\cite{Ref12}, and topological insulators such
as Bi2Se3~\cite{Ref13}.\\
Thus far, the applications of the Dirac equation and relativistic quantum have been addressed
in different fields of physics, including in CMP. However, some of the CMP such as graphene can be considered as cost-effective laboratories to test some of the
relativistic quantum phenomena, including testing the Klein paradox.\\
After an investigation of the importance of the Dirac equation, it is time to address the aim of
the present article. 
In 1926, Erwin Schrodinger introduced his famous wave equation named "Schrodinger's
wave equation" that was associated with many achievements in the quantum field, however,
there was an important point, that is, this equation was not invariant under Lorentz
transformations. Hence, the first quantum relativistic equation was proposed by Klein-
Gordon to apply to spin-zero particles, but, in 1928, Paul Dirac as one of the founders of
quantum mechanics and quantum electrodynamics proposed his relativistic wave equation for
the electron to solve the negative probability problem in the Klein-Gordon equation~\cite{Ref14}.\\ 
The Dirac equation for a free particle or particles
with constant potential is in the form of $ \left( i\gamma^{\mu}\partial_{\mu}\mp m\right) \Psi_{\textbf{k}}\left( \textbf{r},t\right) =0 $ in the natural units $ \left( \hbar=c=1\right)  $ (as
usual the minus (-) sign is used). As in most relativistic quantum books, including~\cite{Ref14}, the $ \gamma^{\mu} $s
matrices that is known as the Dirac matrix or Dirac representation, is a matrix with even dimension $ \left( 2\times2 \ or \ 4\times4 \ or \ \cdots \right) $ that
is not unique but it should follow some given conditions.\\
Nevertheless, ever since the discovery of the Dirac equation, only a few specific and useful
answer was considered, such as "Standard or Dirac-Pauli Representation" (that we show this with S-index),"Super Symmetric Representation", "Weyl or Spinor representation" and "Majorana representation" here. Up to now, no
one has tried to determine all the acceptable answers for $ \gamma^{\mu} $s matrix. In this work, the $ \gamma^{\mu} $ matrices $\left( 2\times2 \right) $ are
"general representation" (we emphasize this with G-index) is determined with all
conditions governing its elements. It is obvious that of answers corresponding to the Dirac
equation, i.e, from $ \Psi_{\textbf{k}}\left( \textbf{r},t\right) $. to the problem for operators of spin, helicity, Lorentz
transformation, and parity undergo some changes. In sum, all the concepts in which the general representation
are generalized will change with new Dirac matrices of $ \gamma^{\mu} $s, although their original
definition does not change, including Dirac Lagrangian, Dirac Hamiltonian, and continuity
equation of probability in Dirac, etc.\\
Accordingly, the application range of the Dirac equation will be broader than before, and in
the case of deviation between the theoretical results and experimental data, answers that are
more consistent may be obtained using the freedom in the elements of $ \gamma^{\mu} $ matrices. It means
one can build unique matrices special for such problems. As Weyl proposed some other
matrices to describe massless relativistic particles with spin-1/2 known as Weyl fermions,
required matrices, and Majorana fermions that are themselves antiparticles\cite{Ref15}, now, one can
determine the correct matrices corresponding to some other problems that follow the Dirac
equation' general principles. This can be associated with changes in the relationship between
the theoretical physics and experimental world.\\
The rest of this paper is structured as follows. Section 2 briefly addresses the history of the
Dirac equation,then the general Dirac matrices are determined for the case
of $ 2 \times 2 $. Section 3 evaluates the
$ \gamma^{\mu} $s matrices and Dirac spinors under the influence of Lorentz transformation and parity.
Finally, the last section corresponds to the results and future work prospects.

\section{Generalized Representation in Two Dimensions}
particle is  where $\mu=0,1,2,...,D$ and D is the dimension of the space. $p^{\mu}=(i\partial/\partial t , -i\partial/\partial x^{i})\Rightarrow p_{\mu}=(i\partial/\partial t , i\partial/\partial x^{i})=i\partial_{\mu} $\
The source of the Klein-Gordon equation's problem is the possibility of negative probability which is due to the second degree time derivative in it. Dirac believed that if it were possible to take the square root of the Klein-Gordon equation so that the second order time derivative becomes the first order time derivative, the problem would be solved. But clearly, the Klein-Gordon equation is not a perfect square. Hence, he assumed that $p^{\mu}p_{\mu}=(\gamma^{\mu}p_{\mu})^{2}=\gamma^{\mu}p_{\mu}\gamma^{\nu}p_{\nu}$ can be written using an unspecified coefficients called the gamma coefficients. Assuming $[p_\mu , \gamma^{\nu}]=0$, then $\gamma^{\mu}\gamma^{\nu}p_\mu p_\nu=p^\mu p_\mu$ has to hold. As we know, it's possible to write $\gamma^{\mu}\gamma^{\nu}p_\mu p_\nu=g^{\mu\nu}p_\mu p_\nu$ using Minkowski metric tensor $(diagg^{\mu\nu}=(1,-1,-1,...) )$. It might seem that by removing the term $p_\mu p_\nu$ from this equation, one can write $\gamma^{\mu}\gamma^{\nu}=g^{\mu\nu}$. But this cannot be done because of the presence of the in dices of $\nu$ and $\mu$ in other terms of this equation. Hence, the symmetric property in Minkowski metric tensor should be considered. Since the $g^{\mu\nu}$ matrix is symmetric, $\gamma^{\mu}\gamma^{\nu}$ has to be written symmetrically as well. After doing some calculations and removing the antisymmetric term, the following fundamental equation is obtained:
\begin{equation}
	\left\lbrace \gamma^{\mu},\gamma^{\nu}\right\rbrace =2g^{\mu\nu}I
\end{equation}
This anti-commute relation confirms that the gamma coefficients cannot be numbers. Hence, Dirac searched for their next from; the gamma matrices. 
According to the Eq. 1, other useful results are obtained which is the focus of this manuscript and help us to know the elements of these matrices better. (In the whole article, the Greek letters start from zero and the English letters from one)\\
I.	The gamma matrices are matrices of even dimension.\\
Proof:
$$
\left\lbrace\gamma^{0},\gamma^{i}\right\rbrace =0\Rightarrow\gamma^{0}\gamma^{i}=-\gamma^{i}\gamma^{0} $$
$$
\Rightarrow
\left|\gamma^{0}\gamma^{i} \right| =\left|-\gamma^{i}\gamma^{0}\right|=(-1)^{N}\left|\gamma^{i}\gamma^{0}\right|
$$
$$
\Rightarrow\left|\gamma^{0}\right|\left|
\gamma^{i} \right|=(-1)^{N}\left|\gamma^{i}\right|\left|\gamma^{0}\right|\Rightarrow N=Odd.
$$
II.	$\left\lbrace \gamma_{\mu},\gamma_{\nu}\right\rbrace  = 2g_{\mu\nu}I=2g^{\mu\nu}I$ . \\
Proof: We know that $[g_{\mu\nu},\gamma^{\lambda}]=[g_{\mu\nu},g_{\theta\varphi}]=0$
$$
\gamma_{\mu}\gamma_{\nu}+\gamma_{\nu}\gamma_{\mu}=g_{\mu\theta}
\gamma^{\theta}g_{\nu\varphi}\gamma^{\varphi}+g_{\nu\varphi}\gamma^{\varphi}
g_{\mu\theta}\gamma^{\theta}$$
$$
=g_{\mu\theta}g_{\nu\varphi}(\gamma^{\theta}\gamma^{\varphi}+\gamma^{\varphi}\gamma^{\theta})=2g_{\mu\theta}g^{\theta\varphi}g_{\nu\varphi}I=2g_{\mu\nu}
I.
$$
III.	If there is an invertible matrix that can transform the gamma matrices to their similar, then the similar matrices follow this fundamental relation $\left\lbrace\gamma'^{\mu},\gamma'^{\nu} \right\rbrace =2g^{\mu\nu}I $.\\ 

Proof: $\gamma'^{\mu}=A\gamma^{\mu}A^{-1},\gamma'^{\nu}=A\gamma^{\nu}A^{-1} $ then
$$
\left\lbrace\gamma'^{\mu},\gamma'^{\nu}\right\rbrace=A\gamma^{\mu}A^{-1}A\gamma^{\nu}A^{-1}+A\gamma^{\nu}A^{-1}A\gamma^{\mu}A^{-1}$$
$$
=A(\gamma^{\mu}\gamma^{\nu}+\gamma^{\nu}\gamma^{\mu})A^{-1}=2g^{\mu\nu}I. $$
IV.The gamma matrices can be diagonalized.	 \\
Proof: if $P^{-1}\gamma^{\mu}P=\gamma^{\mu}_{D}$, that index-D means diagonal matrix. And we saw $(\gamma^{\mu})^{2}=\pm I$ so
$$
(\gamma^{\mu}_{D})^{2}=P^{-1}\gamma^{\mu}PP^{-1}\gamma^{\mu}P
=P^{-1}(\gamma^{\mu})^{2}P=\pm I.$$
V.	$(\gamma^{0}_{D})^{2}=I$.\\
Proof:If $P^{-1}\gamma^{0}P=\gamma^{0}_{D}$ and $I=P^{-1}IP=P^{-1}(\gamma^{0})^{2}P=P^{-1}P\gamma^{0}_{D}P^{-1}P\gamma^{0}_{D}P^{-1}P=(\gamma^{0}_{D})^{2}$.\\
VI.	The eigenvalues of the matrix $\gamma^{0}$ is $\pm1$.\\
Proof:
$I=(\gamma^{0}_{D})^{2}=
\begin{pmatrix}
\lambda_{1}^{2} & 0 & \cdots \\
0 & \lambda_{1}^{2} & \cdots \\
\colon &\colon &\ddots
\end{pmatrix}\Rightarrow \lambda_{i}=\pm1
$. \\
VII.$(\gamma^{i}_{D})^{2}=-I$.\\	
Proof: $P^{-1}\gamma^{i}P=\gamma^{i}_{D}$ and $-I=-P^{-1}IP=P^{-1}(\gamma^{i})^{2}P=P^{-1}P\gamma^{i}_{D}P^{-1}P\gamma^{i}_{D}P^{-1}P=(\gamma^{i}_{D})^{2}$.\\
VIII.	The eigenvalues of the matrix $\gamma^{i}$ is $\pm i$.
Proof: $-I=(\gamma^{i}_{D})^{2}=
-\begin{pmatrix}
\lambda_{1}^{2} & 0 & \cdots \\
0 & \lambda_{1}^{2} & \cdots \\
\colon &\colon &\ddots
\end{pmatrix}\Rightarrow \lambda_{i}=\pm i.
$ \\
IX.	The gamma matrices are traceless. \\
Proof: We know that $tr(abc)=tr(cab)=tr(bca)$
$$
if \ \ \mu\neq\nu\Rightarrow \gamma^{\mu}\left\lbrace\gamma^{\mu},\gamma^{\nu} \right\rbrace=0 $$ 
$$
\Rightarrow\pm\gamma^{\nu}=-\gamma^{\mu}\gamma^{\nu}\gamma^{\mu}\Rightarrow tr(\pm\gamma^{\nu})$$
$$=-tr(\gamma^{\mu}\gamma^{\nu}\gamma^{\mu}) 
=-tr(\pm\gamma^{\nu})\Rightarrow tr(\gamma^{\nu})=0. \ 
$$
X.	$\gamma^{\mu}\gamma^{\nu}\neq 0$.\\
Proof:
$$
|\gamma^{\mu}\gamma^{\nu}|=|A \gamma^{\mu}_{D}A^{-1}B\gamma^{\mu}_{D}B^{-1}|=|\gamma^{\mu}_{D}\gamma^{\mu}_{D}|\neq0\Rightarrow\gamma^{\mu}\gamma^{\nu}\neq 0.
$$
XI.	The gamma matrices are normal which means $[\gamma^{\mu},(\gamma^{\mu})^{\dagger}]=0$.\\
Proof: Because $\gamma^{\mu}=A \gamma^{\mu}_{D}A^{-1}$ and $A$ is orthonormal.\\
XII.	$(\gamma^{0})^{\dagger}=\gamma^{0}$.\\
Proof: We know that $(\gamma^{0}_{D})\dagger=\gamma^{0}_{D}$ then $(\gamma^{0})\dagger=A^{\dagger}\gamma^{0}_{D}A=\gamma^{0}$.\\
XIII.		$(\gamma^{i})^{\dagger}=-\gamma^{i}$.\\
Proof:  We know that $(\gamma^{i}_{D})\dagger=-\gamma^{i}_{D}$ then $(\gamma^{i})\dagger=A^{\dagger}(\gamma^{i}_{D})^{\dagger}A=-A^{\dagger}\gamma^{i}_{D}A=-\gamma^{i}$.\\
As we know, the Dirac equation is a D+1 dimension equation; meaning that time is always considered as one dimension in this equation. On the other hand, each of the gamma matrices corresponds to a dimension of space-time. Hence,in which is only one number,is assigned to the dimension of time;and since j=1,2,...,D; D is assigned to the dimension of space. 
Now, lets consider the smallest matrix; $2\times2$ gamma matrix.
$$
\gamma^{0}=
\begin{pmatrix}
c_{0} & a_{0}-ib_{0} \\
a_{0}+ib_{0} & -c_{0}
\end{pmatrix},$$
$$
\gamma^{1}=i
\begin{pmatrix}
c_{1} & a_{1}-ib_{1} \\
a_{1}+ib_{1} & -c_{1}
\end{pmatrix}, $$
$$
\gamma^{2}=i
\begin{pmatrix}
c_{2} & a_{2}-ib_{2} \\
a_{2}+ib_{2} & -c_{2}
\end{pmatrix}
$$
Where
\begin{align}
	a_{0}=c_{1}b_{2}-b_{1}c_{2}\ ,\ b_{0}=a_{1}c_{2}-c_{1}a_{2}\ ,\ c_{0}=b_{1}a_{2}-a_{1}b_{2}\notag \\
	a_{1}=b_{0}c_{2}-c_{0}b_{2}\ ,\ b_{1}=c_{0}a_{2}-a_{0}c_{2}\ ,\ c_{1}=a_{0}b_{2}-b_{0}a_{2}\notag \\
	a_{2}=c_{0}b_{1}-b_{0}c_{1}\ ,\ b_{2}=a_{0}c_{1}-c_{0}a_{1}\ ,\ c_{2}=b_{0}a_{1}-a_{0}b_{1} 	
\end{align}
According to the mentioned properties of I to XIII, the following relations among the elements of the gamma matrices are obtained:
$$
a_{0}^{2}+b_{0}^{2}+c_{0}^{2}=a_{1}^{2}+b_{1}^{2}+c_{1}^{2}=a_{2}^{2}+b_{2}^{2}+c_{2}^{2}=1\hspace{4.9cm}
$$
$$ 
a_{0}^{2}+a_{1}^{2}+a_{2}^{2}=b_{0}^{2}+b_{1}^{2}+b_{2}^{2}=c_{0}^{2}+c_{1}^{2}+c_{2}^{2}=1 \hspace{4.9cm} $$
$$
a_{0}a_{1}+	b_{0}b_{1}+	c_{0}c_{1}= a_{0}a_{2}+	b_{0}b_{2}+	c_{0}c_{2}\hspace{4.8cm}$$
$$= a_{1}a_{2}+	b_{1}b_{2}+	c_{1}c_{2}=0\hspace{4.9cm}$$
$$ 
a_{0}b_{0}+	a_{1}b_{1}+	a_{2}b_{2}=	a_{0}c_{0}+	a_{1}c_{1}+	a_{2}c_{2}\hspace{4.9cm}$$
$$=b_{0}c_{0}+	b_{1}c_{1}+	b_{2}c_{2}=0\hspace{4.9cm}
$$
Which can be summarized using the Einstein summation convention as the following: 
\begin{align}
	a_{\mu}=-b_{\nu}c_{\theta}\in_{\mu\nu\theta}\ ,\ b_{\mu}=-c_{\nu}a_{\theta}\in_{\mu\nu\theta}\ ,\ c_{\mu}=-a_{\nu}b_{\theta}\in_{\mu\nu\theta}\notag \\
	a_{\mu}	a_{\mu}=b_{\mu}	b_{\mu}=c_{\mu}c_{\mu}	=a_{\mu}	b_{\mu}=a_{\mu}	c_{\mu}=b_{\mu}c_{\mu}=1,\ \ \ \ \ \	
	\notag \\	a_{\mu}	a_{\nu}+b_{\mu}	b_{\nu}	+c_{\mu}c_{\nu}	=\delta_{\mu\nu},\ \ \ \ \ \ \ \ \ \ \ \ \ \ \ \ \ \ \ \ \
\end{align}
Furthermore, the following relations hold among the gamma matrices themselves:
\begin{align}
	\gamma^{0}\gamma^{1}=-i\gamma^{2},	\gamma^{1}\gamma^{2}=+i\gamma^{0},	\gamma^{2}\gamma^{0}=-i\gamma^{1},-i\gamma^{0}\gamma^{1}\gamma^{2}=I
\end{align}
All properties of the elements of the gamma matrices can be summarized in the following normal matrix:
\begin{align}
	A\equiv 
	\begin{pmatrix}
		c_{0} & c_{1} & c_{2} \\
		b_{0} & b_{1} & b_{2} \\
		a_{0} & a_{1} & a_{2}
	\end{pmatrix};AA^{T} =A^{T}A=I, |A|=1
\end{align}
Hence, matrix A is a member of the rotation group of SO(3). According to the matrices algebra, it can be proved that no other $2\times2$ gamma matrix could be found that is linearly independent of the other two matrices. In other words, no fourth matrix could be found that hold all the mentioned 13 properties along with other three matrices simultaneously. Hence, it can be concluded that this $2\times2$ matrix is at best useful for the 2+1 dimension space. It worth mentioning that if we assume  $c_{0}=b_{1}=a_{2}=1,Etc=0,$ then the normal or standard representation of Dirac equation is obtained; 
\begin{align}
	\gamma^{0}=
	\begin{pmatrix}
		1 & 0 \\
		0 & -1
	\end{pmatrix},
	\gamma^{1}=
	\begin{pmatrix}
		0 & 1 \\
		-1 & 0
	\end{pmatrix},
	\gamma^{2}=i
	\begin{pmatrix}
		0 & 1 \\
		1 & 0
	\end{pmatrix}, 
\end{align}

\subsection{The general form of Dirac equation and its spinors}
In this subsection, the general form of Dirac equation and its solutions are discussed in addition to comparing them to the standard representation of Dirac equation. 
Based on what was stated, the non-differential form of Dirac equation for the 2+1 dimension space has to be as the following:
\begin{table*}
	\begin{align}
		(\gamma^{0}E-\gamma^{1}k_{1}-\gamma^{2}k_{2}-mI)u(E,\vec{k})e^{i(k_{1}x+k_{2}y-Et)}=0
		\notag \ \ \ \ \ \ \ \ \ \ \ \ \ \ \ \ \  \ \ \ \ \ \ \ \ \ \ \ \ \ \ \ \ \  \ \  
		\\{\small \Rightarrow
			\begin{pmatrix}
				c_{0}E -i c_{1}k_{1} -i c_{2}k_{2}-m & (a_{0}-ib_{0})E-i(a_{1}-ib_{1})k_{1}-i(a_{2}-ib_{2})k_{2} \\
				(a_{0}-ib_{0})E-i(a_{1}-ib_{1})k_{1}-i(a_{2}-ib_{2})k_{2} &	-c_{0}E +i c_{1}k_{1} +i c_{2}k_{2}-m
			\end{pmatrix}
			\begin{pmatrix}
				u_{1} \\ u_{2}
			\end{pmatrix}=0}
	\end{align}
\end{table*}

Which is written as follows in the Standard Representation of Dirac equation:
\begin{align}
	\begin{pmatrix}
		E-m & -(k_{1}+ik_{2}) \\
		k_{1}-ik_{2} & -(E+m)
	\end{pmatrix}
	\begin{pmatrix}
		u_{1S}\\
		u_{2S}
	\end{pmatrix}=0 
\end{align}
The determinant of this matrix has to be zero so that the Eq. (9) could have non-trivial solutions. Hence, the important and obvious result $E=\pm\sqrt{k_{1}^{2}+k_{2}^{2}+m^{2}}$ is obtained. 
The solutions of the Equation (9) for two given eigenvalues of $\pm E$ is respectively 
\begin{align}
	u_{+}=N
	\begin{pmatrix}
		(c_{0}E+m)-i(c_{1}k_{1}+c_{2}k_{2})	
		\\
		i(b_{0}E-a_{1}k_{1}-a_{2}k_{2})+(a_{0}E+b_{1}k_{1}+b_{2}k_{2})
	\end{pmatrix}\notag \\
	u_{-}=N
	\begin{pmatrix}
		i(b_{0}E-a_{1}k_{1}-a_{2}k_{2})-(a_{0}E+b_{1}k_{1}+b_{2}k_{2})	
		\\
		(c_{0}E+m)+i(c_{1}k_{1}+c_{2}k_{2})
	\end{pmatrix}\notag \\	
	N=\dfrac{1}{\sqrt{2E(E+c_{0}m+c_{1}k_{2}-c_{2}k_{1})}} \ \ \ \ \ \ \ \ \ \ \ \ \ 		
\end{align}
Which in the standard representation are reduced to the known following solutions:
\begin{align}
	u_{S+}=\dfrac{1}{\sqrt{2E(E+m)}}
	\begin{pmatrix}
		E+m	
		\\
		k_{1}-ik_{2}
	\end{pmatrix}\notag \ \ \ \ \\
	u_{S-}=\dfrac{1}{\sqrt{2E(E+m)}}
	\begin{pmatrix}
		-(k_{1}+ik_{2})
		\\
		E+m
	\end{pmatrix}	
\end{align} 
\section{the gamma matrices and the Dirac spinors under the Lorentz transformation and parity}
If $ \Lambda $ is Lorentz transformation $ (x'=\Lambda x) $ then 
\begin{equation}
	\begin{cases}
		\psi '(x')=S(\Lambda)\psi(x)\Rightarrow\psi(x)
		=S^{-1}(\Lambda)\psi '(x')\\
		\bar{\psi}
		=\psi^{\dag}\gamma^{0}\Rightarrow
		\bar{\psi}'(x')=\bar{\psi}
		(x)S^{-1}(\Lambda)
	\end{cases}
\end{equation}
the prime means inertial system $ O' $ that moves with velocity $ v/c $ relative to the system $ O $, $ (m'=m,(\gamma^{\mu})'=\gamma^{\mu}) $\\
\begin{align}
	O\rightarrow\left( i\gamma^{\mu}\partial_{\mu}-m\right)\psi(x)=0 
	,\notag \\
	O'\rightarrow\left( i\gamma^{\mu}\partial_{\mu}'-m\right)\psi'(x')=0 \notag \\
	\Rightarrow i\gamma^{\mu}\partial_{\mu}'\psi'(x') -m\psi'(x')=0.
\end{align}
$ \partial_{\mu} $ is a vector because it converts like a vector under the Lorentz transformation ie:
\begin{align}
	\partial_{\mu}=\frac{\partial}{\partial x^{\mu}} =\frac{\partial x '^{\nu}}{\partial x^{\mu}}\frac{\partial}{\partial x '^{\nu}}=\Lambda^{\nu}_{\mu}\partial '_{\nu}, \ \ \ \ \ \ \ \ \ \notag \\
	(13,14)\Rightarrow \left( i\gamma^{\mu}\Lambda^{\nu}_{\mu}\partial '_{\nu}-m\right) S^{-1}(\Lambda)\psi '(x')=0 \notag \\
	\Rightarrow i\gamma^{\mu}\Lambda^{\nu}_{\mu}\partial '_{\nu}S^{-1}\psi '(x')-mS^{-1}\psi '(x')=0  \ \ .
\end{align}
We multiply $ S(\Lambda) $ at the left on (15) until the second term of this relation equals to the second term of relation (14):
\begin{equation}
	iS\gamma^{\mu}\Lambda_{\mu}^{\nu}\partial ' _{\nu}S^{-1}\psi '(x')-m\psi '(x')=0
\end{equation}
$ \partial ' _{\nu}S^{-1}=S^{-1}\partial ' _{\nu} $ because the elements S and S-1 are constants and the differential operator $ \partial ' $ or $ \partial $ passes through them. By using (14), (15):
\begin{equation}  
	iS\gamma^{\mu}\Lambda_{\mu}^{\nu}S^{-1}\partial ' _{\nu}\psi '(x')=i\gamma^{\nu}\partial_{\nu}'\psi'(x'),
\end{equation}
then
\begin{equation}
	\gamma^{\nu}=S\gamma^{\mu}\Lambda_{\mu}^{\nu}S^{-1}\longleftrightarrow S^{-1}\gamma^{\nu}S=\gamma^{\mu}\Lambda_{\mu}^{\nu}.
\end{equation}
In the 1+1-dimensional space we have just one "boost" $ (v=tanh\theta) $
\begin{equation}
	\begin{cases}
		x'^{0}=(cosh\theta)x^{0}-(sinh\theta)x^{1}, \\ x'^{1}=(cosh\theta)x^{1}-(sinh\theta)x^{0} 
	\end{cases}
\end{equation}
\begin{equation}
	\Rightarrow
	\begin{cases}
		\Lambda_{0}^{0}=\Lambda_{1}^{1}=cosh\theta \ , \  \Lambda_{0}^{1}=\Lambda_{1}^{0}=-sinh\theta \\
		S=exp(-\frac{\theta}{2}\gamma^{0}\gamma^{1})
		=exp(-\frac{\theta}{2}\bar{\gamma}^{2})=Icosh\frac{\theta}{2}-\bar{\gamma}^{2}sinh\frac{\theta}{2}.
	\end{cases}
\end{equation}
Thus for GR
\begin{equation}
	S=
	\begin{pmatrix}
		cosh \theta/2 \ -c_{2}sinh \theta/2 & -(a_{2}-ib_{2})sinh \theta/2 \\
		-(a_{2}+ib_{2})sinh \theta/2 & cosh \theta/2 \  +c_{2}sinh \theta/2
	\end{pmatrix}.
\end{equation}
Similarly, it can prove that the following relation is also hold:
\begin{equation}
	S^{-1}\gamma^{1}S=\gamma^{0}\Lambda_{0}^{1}+
	\gamma^{1}\Lambda_{1}^{1}=-\gamma^{0} sinh\theta+\gamma^{1}cosh\theta
	.\end{equation}
Also we know the Dirac spinors under the Lorentz transformations will be as follows:
\begin{equation}
	\psi'(x')=S(\Lambda)\psi(x) \ \ or \ \ u'_{\pm}=Su_{\pm}
	,\end{equation}

\subsection{The parity operator}
\label{sec:3}

The parity operator (P) is a special case of Lorentz transformation, namely S, therefore $ \Lambda_{\mu}^{\nu}\gamma^{\mu}=P\gamma^{\nu}P^{-1} $.
in (1+1)-dimensional
\begin{equation}
	\begin{cases}
		x'^{0}=x^{0} \\ 
		x'^{1}=-x^{1}
	\end{cases}
	\Rightarrow 
	\Lambda=
	\begin{pmatrix}
		1 & 0 \\ 0 & -1
	\end{pmatrix},
\end{equation}
then in GR, $ P=e^{i\varphi}\gamma^{0}=e^{i\varphi}
\begin{pmatrix}
c_{0} & a_{0}-ib_{0} \\ a_{0}+ib_{0} & -c_{0}
\end{pmatrix} $. Assuming
\begin{align}
	u_{+}=N
	\begin{pmatrix}
		1 \\ u_{1}
	\end{pmatrix} \ , \  
	u_{-}=N
	\begin{pmatrix}
		-u_{1}^{*} \\ 1
	\end{pmatrix} \ \ \ \ \ \ \ \ \ \ \ \ \ \
	\\
	\Rightarrow
	\begin{cases}
		Pu_{+}=N'e^{i\varphi}
		\begin{pmatrix}
			1 \\ \dfrac{[c_{0}u_{1}-(a_{0}+ib_{0})](-u_{1})}
			{c_{0}u_{1}+(a_{0}-ib_{0})u_{1}^{2}}
		\end{pmatrix} \\ \\
		Pu_{-}=N''e^{i\varphi}
		\begin{pmatrix}
			\dfrac{[c_{0}u_{1}^{*}-(a_{0}-ib_{0})](-u_{1}^{*})}
			{c_{0}u_{1}^{*}+(a_{0}+ib_{0})u_{1}^{*2}} \\ -1
		\end{pmatrix}
	\end{cases},
\end{align}
such that $ N' $ and $ N'' $ are normalization constants.\\
By changing the internal elements of gammas, the definitions that we have for scalar (such as $ \bar{\psi}\psi=\psi\dag\gamma^{0}\psi $) and pseudo-scalar and vector and pseudo-vector don't change because, in spite of using the gammas in these definitions, their reaction to the Lorentz transformation that shows the type of being scalar or vector of them, does not related to the internal elements of gammas.\\
Altogether, concepts in which the Dirac gamma matrices are used, will generalized because of generalizing these matrices, however, the fundamental definition of them does not change such as the Dirac Lagrangian, the Dirac Hamiltonian, the probability continuity equation in Dirac, etc.

\section{Summery}
\label{sec:4}
It was shown in Sec. 2 that how the fundamental Eq. (1) could result in the general representation of Dirac equation (9) as well as other useful relations like the properties of the gamma matrices (I-XIII) in addition to the relation among the elements of the gamma matrices (7). After finding the general form of the Lorentz transform operator, $S(\Lambda)$, in Sec. 3, its effects on the general Dirac spinors were studied. And finally, the general form of the parity operator and its effects on Dirac spinors were studied. The main novelty of this manuscript which makes it important is that despite of all works done in the past nine decades for proving the multiplicity of the representation of Dirac matrices, no efforts have been done in the books and papers to introduce the general form of Dirac matrices. Instead, one specific representation of Dirac matrices is used according to the requirement of the study;such as the standard representation known as Dirac-Pauli Representation, or some especial forms of Supersymmetric Representation or Weyl Representation or Majorana Representation. Sometimes even different forms of Weyl Representation or Majorana Representation are introduced in the papers without mentioning their common origin. Finding the general form of this equation in the present manuscript (for 2+1 dimension) has opened a new way to quick develop of this well known and useful equation in all fields of Physics. It might also lead to the opening of new windows on the influence of Dirac equation; such as the discovery of new particles correspond to its new representation or the discovery of new symmetries in theoretical Physics.
It worth mentioning that there are many works to be done for developing the general form of Dirac matrices;such as finding the general form of Dirac matrices for 3+1 dimension of space-time,as well as greater dimensions. In addition to the discovery of different applications of this form of representation; such as calculating the reflection and transmission coefficients of different particles.And especially, studying Klein tunneling using this representation that is been doing by the authors of the present manuscript right now.

\bibliography{biblio}

\end{document}